\documentclass{WileyMSP-template}
\usepackage{amsmath}
\usepackage{amssymb}
\usepackage{cite}
\usepackage{nicefrac,xfrac}
\newcommand\norm[1]{\left\lVert#1\right\rVert}
\usepackage{ragged2e}
\justifying
\begin{document}

\pagestyle{fancy}
\rhead{\includegraphics[width=2.5cm]{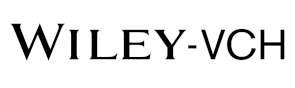}}

\title{Designing variational ansatz for quantum-enabled simulation of non-unitary dynamical evolution- an excursion into Dicke supperradiance}

\maketitle


\author{Saurabh Shivpuje}
\author{Manas Sajjan}
\author{Yuchen Wang}
\author{Zixuan Hu}
\author{Sabre Kais\thanks{kais@purdue.edu}}



\begin{affiliations}
Saurabh Shivpuje, Dr. Manas Sajjan, Dr. Yuchen Wang, Dr. Zixuan Hu, Prof. Sabre Kais\\
Department of Chemistry, Department of Physics and Purdue Quantum Science and Engineering Institute, Purdue University, West Lafayette, Indiana 47907, USA\\
Email Address: kais@purdue.edu


\end{affiliations}



\begin{abstract}
Adaptive Variational Quantum Dynamics (AVQD) algorithms offer a promising approach to providing quantum-enabled solutions for systems treated within the purview of open quantum dynamical evolution. In this study, we employ the unrestricted-vectorization variant of AVQD to simulate and benchmark various non-unitarily evolving systems. We exemplify how construction of an expressible ansatz unitary and the associated operator pool can be implemented to analyze examples such as the Fenna–Matthews–Olson complex (FMO) and even the permutational invariant Dicke model of quantum optics. We furthermore show an efficient decomposition scheme for the ansatz used, which can extend its applications to a wide range of other open quantum system scenarios in near future. In all cases the results obtained are in excellent agreement with exact numerical computations which bolsters the effectiveness of this technique. Our successful demonstrations pave the way for utilizing this adaptive variational technique to study complex systems in chemistry and physics, like light-harvesting devices, thermal, and opto-mechanical switches, to name a few.

\end{abstract}
\section{Introduction}
\label{sec:intro}
Simulation of quantum systems interacting with an inaccessible environment has emerged to be a challenging frontier of research in physics and chemistry not only to demystify its abstruse theoretical underpinning \cite{breuer2002theory,kais2014introduction,hu2020quantum,wang2023simulating} but also to unveil its potential technological ramifications in both natural \cite{sension2007quantum,hu2022general,yeh2012population} and artificial light-harvesting devices \cite{hu2018double,you2017plasmon,candiotto2017charge,hu2021vibration,hu2018dark,roy2021enhancement},  thermal, opto-mechanical switches and transistors \cite{10.1063/1.5096173, PhysRevB.101.184510, PhysRevX.4.011015}, trapping of neutral atoms/ions in a dissipative cavity \cite{PhysRevLett.129.063601}, understanding loss of atomic density in an ultracold lattice \cite{braaten2017lindblad} transport characteristics in atomtronic diodes or transistors\cite{PhysRevA.82.013640}, understanding and mitigating noise on quantum devices \cite{smart2022relaxation,brand2024markovian} to name a few.
The primary challenge in the domain is to accurately represent and simulate the effect of the environment in inducing decoherence and dissipation within the dynamical evolution of the system of interest. Depending on the specific nature of the system and environment, this dynamical evolution can indeed be modelled as memoryless or Markovian \cite{gorini1978properties,sweke2015universal} which assumes an instant recovery of the environment to its initial equilibrium state or it could even be Non-Markovian where once the environment is altered there is a significant back-transfer of information/energy from the latter to the system of interest thereby necessitating the inclusion of memory \cite{head2021capturing}. Such formulations construct generators on a reduced dimension thereby obviating the need to include all the degrees of freedom of the environment which is often intractable. Of the two mentioned, Markovian evolution is the focus of this work due to its inherent simplicity yet pervasive applicability in understanding many intriguing physical phenomena under weak system-environment coupling\cite{sweke2015universal,attal2006open}. A typically employed candidate to model the time dependent trajectory of the system under such Born-Markov approximations \cite{merkli2022dynamics} is the Gorini-Kossakowski-Sudarshan-Lindblad (GKSL) equation, also known as the Lindblad equation \cite{lindblad1976generators,kossakowski1972quantum,gorini1978properties}. Since the aforesaid evolution is time-local, it may apparently seem that solving dynamical evolution generated by it may be easily classically simulatable. However, it must be emphasized that for large system sizes, even classically storing such a quantum state can be resource intensive let alone manipulation. Thus advances made in the direction of constructing new algorithms which are polynomially efficient in run-time and storage would be beneficial.

Over the past few decades, a scientific quest to engender a second quantum revolution have gathered momentum. At the core of this paradigmatic transformation, lies the ability to successfully construct quantum computers and the concomitant
development of a plethora of algorithms which can leverage its unprecedented power~\cite{arute2019quantum,lund2017quantum,sajjan2022quantum,sajjan2023imaginary,li2023toward,shor1999polynomial,aspuru2005simulated}. It is needless to mention that physical sciences stand as an eminent beneficiary due to the innate ability of quantum devices to better simulate quantum mechanical problems itself owing to clever usage of the power of superposition, interference and non-classical correlations. As a result several proposals have been made in recent years which focuses on dynamical evolution of both closed and open systems \cite{hu2020quantum,han2021experimental,wang2011quantum,su2020quantum,endo2020variational,suri2023two,yao2021adaptive, zhang2023quantum, benedetti2021hardware}. In the interest of the latter domain, a primary difficulty lies in the very nature of the evolution itself which demands non-unitary operations unlike the majority of quantum gates which are inherently designed to be reversible and unitary. Most of the algorithms proposed circumvents this restriction through techniques like dilation \cite{hu2020quantum}, singular value decomposition \cite{schlimgen2021quantum}, quantum singular value transformations \cite{suri2023two}, Trotterization of the Liouvillian \cite{PhysRevLett.127.020504}, annealing \cite{chen2022hamiltonian} and even full simulation of the environment and system in an optimized Hilbert space that can faithfully reproduce environment correlation functions\cite{su2020quantum}. Most of these techniques require large-depth circuits which are not near-term implementable as present day quantum devices are prone to imperfect gate operations and short lifetime of the qubits which severely limits the quality and scope of the simulations performed. To this end, variational algorithms have been used with astounding success in the past few years to tackle these problems. One such promising variational method in the domain of open quantum system simulation is the recently proposed Unrestricted Adaptive Variational Quantum Dynamic (UAVQD) method ~\cite{chen2024adaptive}. The crux of the idea is to cast the Lindblad equation into a form that can be interpreted as an effective Schrödinger equation for a non-Hermitian Hamiltonian. The unitary part of the evolution of the quantum state under the effect of the re-formulated equation is thereafter modelled on a quantum device using a variational ansatz. The loss of normalization of the state at each time instant is accounted through a measurement protocol. An unrestricted adaptive scheme is adopted which keeps the norm difference between the actual and the simulated evolution below a preset threshold. UAVQD is especially beneficial for its computational efficiency in obtaining the effective Hamiltonian and its scalability advantages, as the ansatz is chosen by the user.


In this article we use the said technique for performing simulations and benchmarking few important yet a diverse class of non-unitarily evolving systems. Specifically we implement the UAVQD for three physical systems. First system we tackle is the amplitude-damping channel that is widely used as a model for spontaneous emission of a 2-level atom~\cite{hu2020quantum} and also for understanding how dissipation in quantum devices hampers qubit lifetime.
The second is the Fenna–Matthews–Olson complex (FMO) with 3 out of the usual 7 chromophores included~\cite{hu2022general}.
The FMO complex is a widely studied model system for explicating light-induced events in certain strains of autotrophic/photosynthetic bacteria \cite{blankenship2021molecular,scholes2017using}. The FMO mediates the transfer of solar energy from the antenna towards the reaction center. The efficiency of this energy transfer is extremely high as compared to artificial photovoltaics: for this reason, a thorough understanding of energy transfer events within the FMO complex may lead to the development of high-efficiency photovoltaic systems. We thereafter shift our focus to the implementation of this scheme to study the Dicke model of quantum optics and exploring the superradiance effects- a study to the best of our knowledge is hitherto unexplored by any quantum algorithm. Dicke super radiance denotes the collective emission of an ensemble of inverted atomic emitters placed within an array of size much smaller than the wavelength of common electromagnetic mode with which they are interacting~\cite{dicke1954coherence}. Under such conditions, for dilute arrays, the spontaneous emission rate decays monotonically in time and at a given instant is simply proportional to the number of emitters. However, when the array is densely packed, it is observed that interaction among the atoms leads to synchronous emission with the phase of oscillations of the individual atomic dipoles being locked as the emission event proceeds in real time thereby enhancing the emission rate initially. This manifests itself as the characteristic bursts of photon and a distinctive increase in intensity of emission at initial times. Even though the original Dicke problem \cite{dicke1954coherence} is well-understood due to permutational symmetry of the atomic emitters which makes it analytically solvable and tractable, however, in recent years, it has been experimentally observed that such superradiance burst is still retained even in extended and disordered arrays (provided inter-atomic separation is small relative to the wavelength of the common mode). In such systems often site-dependent dipole-dipole interaction between the atomic emitters exist thereby moving away from the aforesaid Dicke limit ~\cite{masson2022universality}.The state space in this regime scales exponentially in the number of emitters as permutational invariance between atoms is lost thereby rendering classical computational efforts hard. We seek to understand this regime in 1D and 2D using the polynomially scaling UAVQD algorithm and efficiently calculate the emission rate. Such efforts may present a new paradigm which may find application in quantum metrology ~\cite{kramer2016optimized}, atomic waveguides ~\cite{asenjo2017exponential} and even in ultracold atoms ~\cite{sierra2022dicke}. 

The organization of this article is as follows. In Section \ref{sec:UAVQD}, we give a review of the UAVQD vectorization method, explaining the mathematical formulations and key steps in the execution procedure. In the next section, Section \ref{sec:appl-examples}, we demonstrate quantum simulations of three open quantum systems: the amplitude damping model, the FMO complex, and Dicke superradiance. For all these examples, we provide a very brief review of their importance and the required theoretical background. We mention the details of the operator pool created for the quantum circuit and simulation parameters. Then we report simulation results obtained for each example in their corresponding subsections in Section \ref{sec:appl-examples}. Finally in Section \ref{sec:conclusion}., we conclude by stating the merits of this method and the scope for future improvements.


\section{Unrestricted
Adaptive Variational Quantum Dynamics}
\label{sec:UAVQD}

The UAVQD approach can potentially emerge as a promising technique for exploring complex quantum phenomena, such as the Dicke superradiance and the Fenna-Matthews-Olson (FMO) complex. This method, as recently discussed by Chen et al.~\cite{chen2024adaptive}, offers a compelling framework for simulating the quantum dynamical evolution of open systems for NISQ devices. AVQD is available in two variants: the quantum trajectory method\cite{yip2018quantum} and the vectorization method\cite{horn1991topics,kamakari2022digital}, with the latter being the central focus of this article.


The primary objective of the vectorization method is to respect the fact that digital quantum computers can strictly implement only unitary operations, yet adapt the latter to simulate non-unitary dynamical evolution. This adaptation is achieved by transforming the density matrix $ \rho \in \mathcal{L}(C^{N})$ as follows: 

\begin{equation}
\rho \mapsto \text{vec}(\rho) = |\nu\rangle \:\:\:\:\: \rm{where} \:\:\:\:
|\nu\rangle \equiv [\rho_{11},\rho_{12},\cdots,\rho_{21},\rho_{22},\cdots,\rho_{NN}]^{\mathrm{T}} \in C^{N^2}
\end{equation}

It is easy to ensure, that the following relation holds 
\begin{equation}
\sqrt{\mathrm{Tr}(\rho^{\dagger}\rho)} = \sqrt{\langle \nu |\nu\rangle},
\end{equation}
Since during any non-unitary dynamical evolution, the purity of the density matrix $\rho$ is lost, normalization of $|\nu\rangle$ is also compromised, a preeminent feature which we shall return to soon. Given, that the system of interest is described by $H \in \mathcal{L(C^N)}$ and jump operators $\{L_k\}_{k=0}^p$ where $L_k \in \mathcal{L(C^N)} \:\: \forall\:k$, the time evolution of \(|\nu\rangle\) can be expressed as :

\begin{equation}\label{dvec_dt}
\frac{d}{dt}|\nu(t)\rangle = - \mathrm{i} H_{\mathrm{eff}}(t)|\nu(t)\rangle,
\end{equation}

In deducing Eq.\ref{dvec_dt} we have re-formulated the standard Gorini–Kossakowski–Sudarshan–Lindblad (GKSL) master equation \cite{lindblad1976generators,kossakowski1972quantum,gorini1978properties}, in Lindbald form using the well-known identity $\text{vec}(ABC)=(C^T\otimes A)\text{vec}(B)$ where $(A,B,C) \in \mathcal{L(C^N)}$. The operator $H_{\mathrm{eff}} \in \mathcal{L}(C^{N^2})$  is an effective non-Hermitian Hamiltonian given by:

\begin{equation}
H_{\mathrm{eff}} = I \otimes H - H^{T} \otimes I + \mathrm{i} \sum_k \left[L_k^{\ast} \otimes L_k - \frac{1}{2}(I \otimes L_k^{\dagger}L_k + L_k^{T}L_k^{\ast} \otimes I ) \right],
\end{equation}

It must be emphasized that $H_{\mathrm{eff}}$ is non-Hermitian and hence can be decomposed into a Hermitian (\(H_e\)) and anti-Hermitian (\(H_a\)) part as follows:

\begin{equation}
    H_{\mathrm{eff}} = H_e - \mathrm{i} H_a,
\end{equation}

where \(H_e = \frac{H_{\mathrm{eff}} + H_{\mathrm{eff}}^{\dagger}}{2}\) and \(H_a = \mathrm{i}\frac{H_{\mathrm{eff}} - H_{\mathrm{eff}}^{\dagger}}{2}\).

To solve the linear ODE in Eq.\ref{dvec_dt} with a non-Hermitian Hamiltonian, we employ a variational quantum algorithm. 
The key concept of a variational algorithm is to approximate quantum states with a parameterized circuit (ansatz):

\begin{equation} \label{generic ansatz}
    |\nu(t)\rangle \approx |\phi(t)\rangle = \prod^k_{l=1} e^{-\mathrm{i}\theta_{l}(t)O_{l}}|\psi_R\rangle,
\end{equation}

where $l$ indexes each layer of the circuit with a unitary gate $e^{-\mathrm{i}\theta_{l}(t)O_{l}} \in \mathcal{L}(C^{N^2})$, \(\theta_{\mu}(t)\) are real tunable parameters, and \(|\psi_R\rangle\) is the initial reference state. The upper limit $k$ in Eq.\ref{generic ansatz} corresponds to the number of operators considered in the pool and can be tuned to enhance the expressibility of the ansatz. The evolution of the state is approximated by the evolution of the parameters \(\theta_{l}(t)\). McLachlan's variational principle is used to minimize the distance as follows:

\begin{equation}\label{Mclaghlan_norm}
    \delta \norm{\frac{d|\phi(\boldsymbol{\theta} (t))\rangle}{dt}+\mathrm{i} H_{\mathrm{eff}}|\phi(\boldsymbol{\theta} (t))\rangle}^2=0.
\end{equation}

Variational minimization as indicated in Eq.\ref{Mclaghlan_norm} leads to the equation of motion for the tunable parameters:

\begin{equation} \label{eq:Mtheta_dot=V}
    \mathbf{M}(t)\dot{\boldsymbol{\theta}}(t)=\mathbf{V}(t),
\end{equation}

where elements of \(\mathbf{M}\) and \(\mathbf{V}\) are defined as 
\begin{eqnarray}\label{eq:M_kj and V_k}
M_{kj}(t) &= 2\rm{Re}(\frac{\partial \langle \phi(\theta(t))|}{\partial \theta_k(t)}\frac{\partial | \phi(\theta(t))\rangle }{\partial \theta_j(t)} + \langle \phi(t)|\frac{\partial | \phi(\theta(t))\rangle }{\partial \theta_k(t)}\langle \phi(t)|\frac{\partial | \phi(\theta(t))\rangle }{\partial \theta_j(t)}) \\
V_k(t)&= 2\rm{Im}(\langle H_{\mathrm{eff}} \rangle \langle \phi(\theta(t))|\frac{\partial | \phi(\theta(t))\rangle }{\partial \theta_k(t)} + \frac{\partial \langle \phi(\theta(t))| }{\partial \theta_k(t)} H_{\mathrm{eff}} |\phi(t) \rangle )
\end{eqnarray}

The form of the ansatz in terms of unitary operations as given in Eq.\ref{generic ansatz} is norm-preserving and hence can only track the evolution of the normalized state. However as indicated before modelling the loss of purity of the density matrix is quintessential. To track the norm loss at a given time $t$ with an increment of 
 \(dt\) one can compute the following:
 \begin{equation}
  \langle\Tilde{\phi}(\theta(t+dt))|\Tilde{\phi}(\theta(t+dt))\rangle\approx e^{-\delta \Gamma}\langle\Tilde{\phi}(\theta(t))|\Tilde{\phi}(\theta(t))\rangle
\end{equation}

where $\delta \Gamma=2\langle\phi(\theta)(t)|H_a|\phi(\theta)(t)\rangle dt$.

The algorithm uses an adaptive procedure to add components to the ansatz i.e. add new operators $\{O_l\}$ from a pool of available operators to enhance expressibility. To ensure that is the case, the McLachlan's distance is kept below a prefixed threshold, and if it exceeds this threshold, the adaptive procedure is triggered. In the unrestricted adaptive variational quantum dynamics (UAVQDS) protocol, a greedy approach is used to select and apply operators from the pool that lower the McLachlan distance, ensuring that it reaches the lowest possible value. 

To simulate a system with an \(\rm{N} \times \rm{N}\) density matrix, one typically needs to read and process \(\rm{N(N-1)}/2\) entries, requiring \(\rm{N(N-1)}\) variables to represent these entries and resulting in a complexity of \(O(\rm{N}^2)\). However in a variational protocol like ours we typically use a circuit construction/ansatz comprising of $2\rm{log}_2 (N)$ qubits and operations characterized by parameterized unitary gates which require $O(\rm{poly}(2log_2 (N))$ parameters in total to represent the target state. Such constructions are made in such a way so that $O(\rm{poly}(2log_2 (N))$ operations/gates (especially two-qubit gates as in near-term devices such operations have high infidelities thereby adversely impacting the quality of the result) are required which affords a polynomial run-time complexity as well. We shall exemplify such ansatz development schemes in the forthcoming sections through appropriate examples. Furthermore the algorithm is equipped with features that allows us to read the matrix elements of ($\rm{M}_{kj}$ of matrix $\mathbf{M}$ and $\rm{V}_k$ of vector $\mathbf{V}$) directly from the quantum circuit through an appropriately designed quantum circuit \cite{endo2020variational}. This can potentially obviate the need to ever access/store the full quantum state which provides a scheme polynomial in system size as far as storage is concerned. Furthermore, the number of measurements required from such quantum circuits would be equal to number of matrix elements of ($\mathbf{M}$, $\mathbf{V}$) both of which inherently depends on the number of parameters used (See Eq.\ref{eq:M_kj and V_k}) and is order $O((\rm{poly}(2log_2 (N)))^2)$. For a more detailed description and implementation of the algorithm, please refer to the original work by Chen et al. in Ref.~\cite{chen2024adaptive}.


\section{Applications}
\label{sec:appl-examples}

\subsection{2-Level Amplitude damping channel}
Before delving into more complex examples, let's examine a simple two-level amplitude damping channel, which is straightforward to implement and can also validate the execution of UAVQD scheme. Amplitude damping channels provide valuable insights into the behavior of quantum systems in realistic environments and are essential for advancing the field of quantum information science. These channels model noise processes, like spontaneous emission or energy dissipation, arising from interactions with the environment. Understanding and mitigating their effects are crucial for designing robust quantum algorithms and implementing fault-tolerant quantum computing schemes.\cite{nielsen2010quantum,wilde2013quantum}, The state space in this example is $2$-dimensional and is labelled as $(|0\rangle, |1\rangle)$. We choose the said levels to be degenerate with the common energy value set to zero. Since the paradigm is famously known for modelling spontaneous emission, we use jump operators of the form $\sigma^{-}=|0\rangle\langle 1|$ facilitating the transition from the state $|1\rangle$ to the  state $|0\rangle$ with a transition rate of $\gamma$, while $\sigma^{+}=(\sigma^{-})^{\dagger}$ is its Hermitian conjugate signifying transition in the reverse process. The equation of motion of the density operator $\rho(t) \in \mathcal{L}(C^2)$ of this system is governed by the usual GKSL equation as:

\begin{equation}
    \frac{d\rho(t)}{dt} = \gamma\left(\sigma^{+}\rho(t)\sigma^{-} -\frac{1}{2}\{\sigma^{-}\sigma^{+},\rho(t)\} \right),
\end{equation}
with $\{\cdot, \cdot \}$ denoting the anti-commutator.

\begin{figure}
  \centering
  \includegraphics[width=0.9\textwidth]{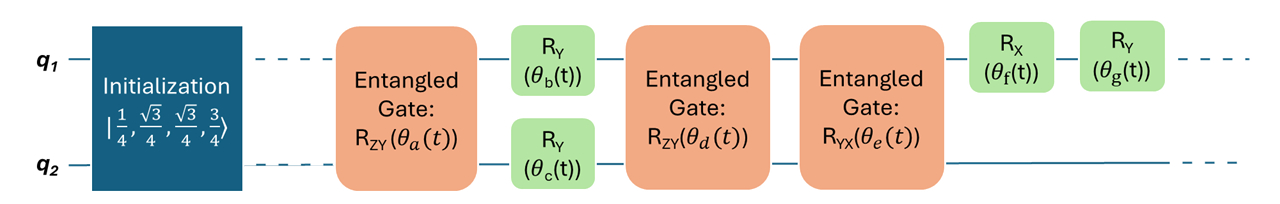}
  \caption{Snippet of the ansatz construction scheme. This figure provides a visual representation of the ansatz used in UAVQD method. The circuit initializes in the state given in first block. Next, the two-qubit entangling operations are labeled as 'Entangled Gate' with notation scheme $R_{ZY}\theta_a(t)$ implies two qubit rotation ZY gate with $\theta_a$ parameter at time t, where subscripts $a,b,c\ldots$ are simply to notify that each block has distinct $\theta$.
} \label{fig:amp_damp_circuit}
\end{figure}

To treat the system, we constructed an ansatz using a pool of operators that included single-qubit Pauli operators $P_{\text{single}} = \{R_{X_i}(\theta(t))\}_{i=1}^2 \cup \{R_{Y_i}(\theta(t))\}_{i=1}^2 \cup \{R_{Z_i}(\theta(t))\}_{i=1}^2$ and two-qubit Pauli operators obtained from all possible combinations of $P_{\text{two}} = e^{\frac{-\mathrm{i}\theta P_i \otimes P_j}{2}}$ where $P_{i}, P_{j} \in \{X_i, Y_i, Z_i\}$. Fig. \ref{fig:amp_damp_circuit} provides a snippet of the ansatz to illustrate the scheme used.
For the numerical simulation, we start with an initial state of $\left[\frac{1}{2}, \frac{\sqrt{3}}{2}\right]$ which corresponds to a $2 \times 2$ density matrix as:
$$
\left[\begin{array}{cc}
0.25 & 0.433013 \\
0.433013 & 0.75
\end{array}\right],
$$

\begin{figure}
  \centering
  \includegraphics[width=0.7\textwidth]{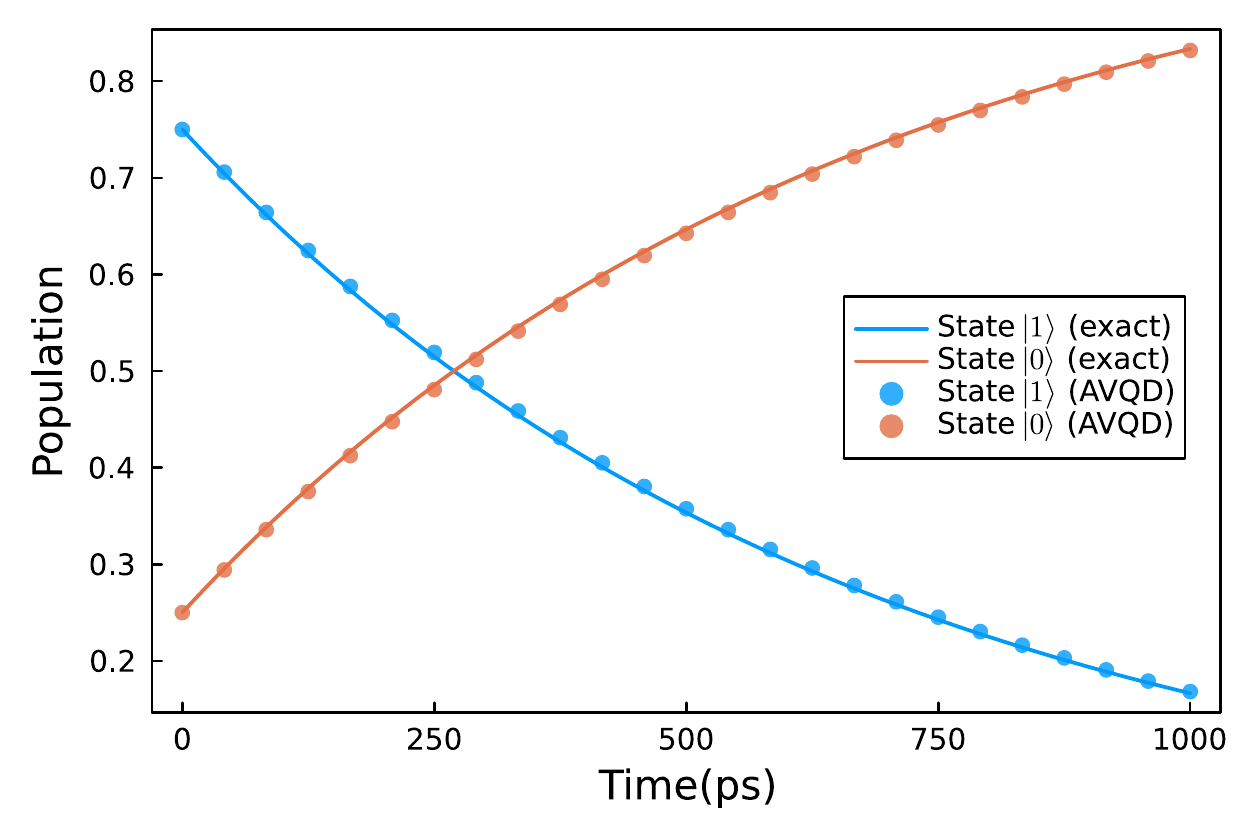}
  \caption{Population comparison between state $|0\rangle$ and state $|1\rangle$ for the amplitude damping model. Solid lines show exact solutions serving as benchmarks, while filled circles represent results from vectorization and UAVQDS methods.}
  \label{fig:amp_damp_data}
\end{figure}

This was then flattened into a $4 \times 1$ vector to initiate the vectorization protocol. We chose a decay rate of $\gamma = 1.52 \times 10^9 \, \text{s}^{-1}$ and performed calculations \cite{hu2020quantum} using the UAVQDS method. The populations of the two states were extracted from the first and fourth entries of the density matrix output at each timestep, shown as filled dots in Fig.\ref{fig:amp_damp_data}. We set an adaptive threshold of $10^{-6}$ for the McLaghlan norm and evolved the system from time $t = 0$ ps to $1000$ ps with a time-step ($\delta t$) of $40$ ps. The solid lines in Fig.\ref{fig:amp_damp_data} depict the exact solution derived from the Hamiltonian Open Quantum System Toolkit (HOQST) package\cite{chen2022hamiltonian}, serving as benchmarks. The UAVQD results (filled dots) align extremely well with the exact solution.
These results demonstrate the accuracy of this method in generating quantum circuit enabled simulations.

\subsection{Fenna-Matthews-Olson (FMO) complex}

To test the efficacy of the UAVQD method for a more complex problem, we chose the Fenna-Matthews-Olson (FMO) complex as an example. FMO is essentially a trimeric pigment protein complex that occurs in green sulfur bacteria and plays a pivotal role in the process of photosynthetis \cite{blankenship2021molecular}. Within each FMO monomer, there are seven bacteriochlorophyll chromophores each of which can be modelled as a pseudospin-$\frac{1}{2}$ system. The initial photo-excitation can arise in either chromophore 1 or 6, and is eventually transferred to chromophore 3, closely linked to the reaction center for the Calvin cycle.\cite{adolphs2006proteins,skochdopole2011functional,bassham1979photosynthetic} This energy transfer involves hopping to neighboring chromophores and predominantly influenced by environmental interactions especially of the protective protein scaffold.\cite{valleau2017absence}

The FMO complex consists of multiple highly efficient routes for exciton transfer to the reaction center. Here, we consider the three-chromophore subsystem, consisting of chromophores 1-3, the ground state, and the sink state, which is known to faithfully replicate the exciton dynamics of the complete seven-chromophore monomer\cite{skochdopole2011functional}. This five site sub-system is labelled as $|i\rangle_{i=0}^{4}$ with excitation at the sink denoted by the state $|4\rangle$ and that of the ground as $|0\rangle$. The Hamiltonian for such a system in this single excitation manifold is given as follows:

\begin{equation}
    H = \sum_{i=0}^4 \omega_i \sigma_i^+ \sigma_i^- + \sum_{j \neq i} J_{ij}(\sigma_i^+\sigma_j^- + \sigma_j^+\sigma_i^-)
    \label{eq:FMO_ham}
\end{equation}
where state $\lvert i \rangle$, characterized by energy $\omega_i$, is generated using the Pauli raising operator $\sigma_i^+$ on the vacuum and eliminated using the Pauli lowering operator $\sigma_i^-$. The coupling strength which defines the transition rate for the coherent dynamics between states $\lvert i \rangle$ and $\lvert j \rangle$ is denoted by $J_{ij}$. The dynamics of the FMO system can be effectively modeled using the GKSL master equation in Lindbald form as.

\begin{equation}
    \frac{d\rho(t)}{dt} = -\mathrm{i}[H,\rho(t)] + \sum_{k>0} \big( L_k\rho(t)L_k^{\dagger} - \frac{1}{2}\{L_k^{\dagger}L_k, \rho(t)\}\big)
    \label{eq:17}
\end{equation}
where the seven operators $L_k$ represent distinct physical processes, each embedding a rate $\gamma_k$ within its definition. Dephasing is described by operators $L_1$ through $L_3$ where $L_{deph}(i) = \sqrt{\alpha} \lvert i \rangle \langle i \rvert$ with $i=1,...,3$; dissipation, captured by operators $L_4$ through $L_6$, each of which depicts the transition from state $\lvert i \rangle$ to the ground state $\lvert 0 \rangle$ as $L_{diss}(i) = \sqrt{\beta} \lvert 0 \rangle \langle i \rvert$ with $i=1,...,3$. Hopping of excitation irreversibly from state $\lvert 3 \rangle$ to the sink ($\lvert 4 \rangle$) is described as $L_{sink} = \sqrt{\gamma} \lvert 4 \rangle \langle 3 \rvert $.

\begin{figure}
  \centering
  \includegraphics[width=0.9\textwidth]{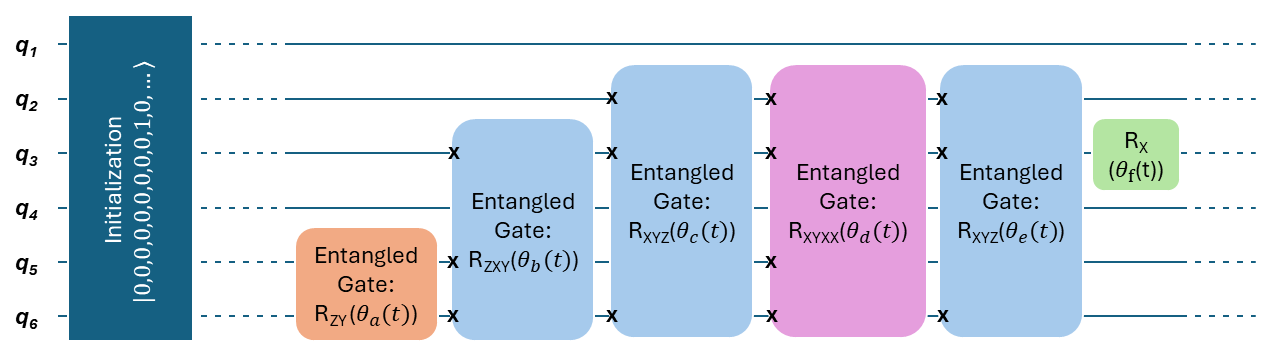}
  \caption{This schematic diagram depicts the part of the quantum circuit utilized in the simulation of the FMO complex, operating with a 6-qubit system and with notation same as the previous Figure \ref{fig:amp_damp_circuit}. It highlights the possible arrangements of single-qubit and multi-qubit entangling operations, where cross marks in the blocks represent the qubits involved in an entanglement.}
  \label{fig:FMO_ansatz}
\end{figure}

Given that the Hamiltonian in Eq.\ref{eq:FMO_ham} is represented by a $5 \times 5$ matrix in the basis of states $|i\rangle_{i=0}^{4}$, it can be padded to represent operators in the space of three qubits. According to the vectorization protocol, which requires twice the number of physical qubits for representation, the ansatz for this system would thus involve six qubits. The operator pool designed for this example is more complex than that used in the amplitude damping model and serves as a precursor for tackling our primary case study of Dicke Superradiance. We have developed a comprehensive pool of operators that incorporates single-qubit Pauli operators \( P_{\text{single}} = \{R_{X_i}\}_{i=1}^6 \cup \{R_{Y_i}\}_{i=1}^6 \cup \{R_{Z_i}\}_{i=1}^6 \), as well as multi-qubit operators ranging from two to four qubits. These are constructed from all possible combinations of P$_{\rm{two}}$ = $e^{i\theta P_i \otimes P_j}$, P$_{\rm{three}}$ = $e^{i\theta P_i \otimes P_j \otimes P_k}$, P$_{\rm{four}}$ = $e^{i\theta P_i \otimes P_j \otimes P_k \otimes P_m}$ where  $\{P_i, P_j, P_k, P_m\} \in \{X_i, Y_j, Z_k\} \forall \:\: (i, j, k) \in \mathcal{Z}_6$ with the condition that \( i \ne j \ne k \). Fig. \ref{fig:FMO_ansatz} provides a visual excerpt of the ansatz, illustrating the utilized scheme.

\noindent To simulate the FMO complex using the UAVQD method, we initialized it with a specific state where only the first chromophore is excited i.e. state $|1\rangle$. Calculations were performed using a time interval of 1 fs until $t_f = 300$ fs. The elements $(\omega_i, J_{ij})$ when substituted in eV in Eq.\ref{eq:FMO_ham}, leads to the following form \cite{hu2022general}

\begin{equation}
H = \begin{pmatrix} 0 & 0 & 0 & 0 & 0\\ 0 & 0.0267 & -0.0129 & 0.000632 & 0\\ 0 & -0.0129 & 0.0273 & 0.00404 & 0\\ 0 & 0.000632 &  0.00404 & 0 & 0\\ 0 & 0 & 0 & 0 &0\end{pmatrix}
\end{equation}

For the numerical solution a dephasing rate $\alpha = 3.00 \times 10^{-3}$ fs$^{-1}$, a dissipation rate $\beta = 5.00 \times 10^{-7}$ fs$^{-1}$, and a sink rate $\gamma = 6.28 \times 10^{-3}$ fs$^{-1}$\cite{hu2022general}. These parameters are crucial for accurately modeling the dynamics of the FMO complex.

\begin{figure}
  \centering
  \includegraphics[width=0.65\textwidth]{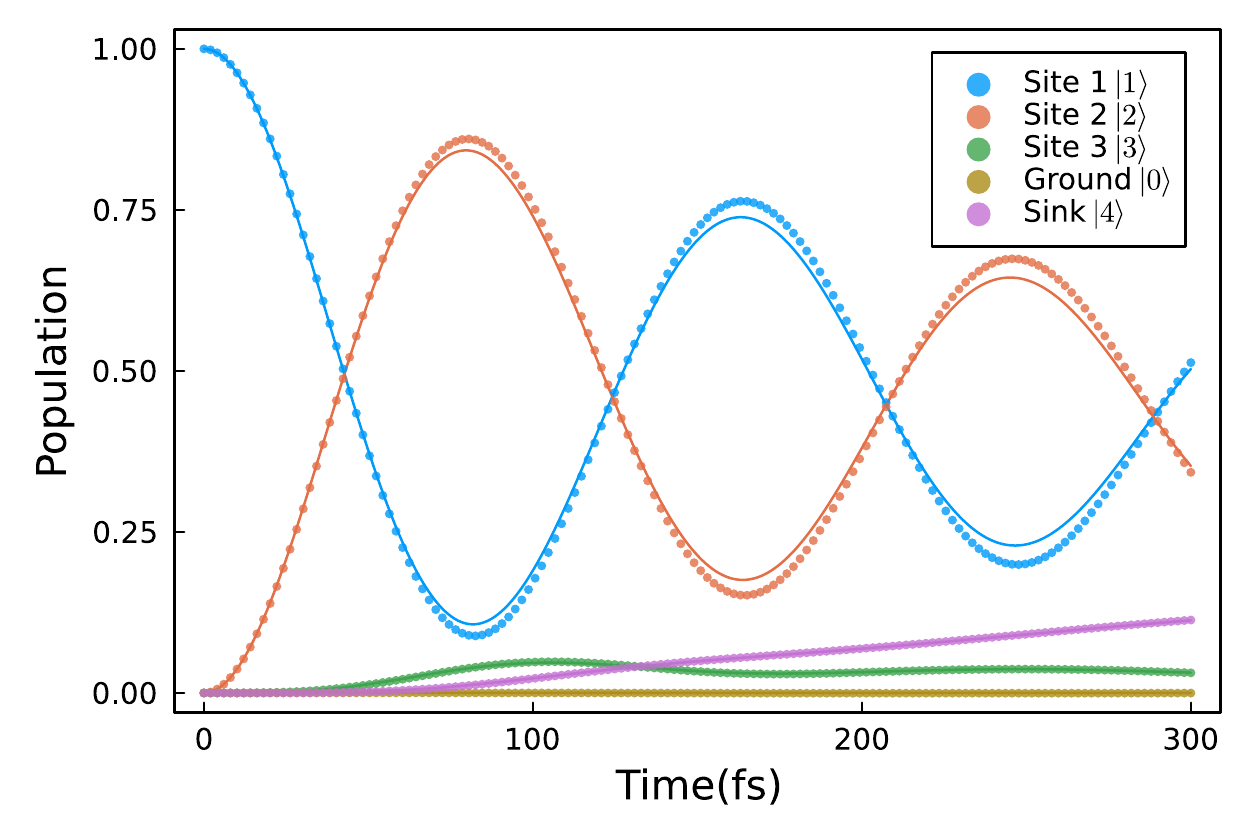}
  \caption{The population dynamics of the Fenna-Matthews-Olson (FMO) complex, showcasing the populations of site 1 through 3, the ground state, and the sink state. The solid lines represent the exact calculations, while the spaced filled dots correspond to calculations using the vectorization and UAVQDS methods. The UAVQD method was employed with a timestep of 1 fs, and an adaptive threshold of $10^{-3}$.}
  \label{fig:FMO_data}
\end{figure}

The results of the UAVQD simulation for the FMO complex were compared to the exact solution, showing very good agreement as can be seen in Fig.\ref{fig:FMO_data}. The populations of the ground state, sink state, and three chromophore sites were extracted from the density matrix output at each timestep, demonstrating the effectiveness of the UAVQD method in accurately simulating the dynamics of systems as complex as the FMO.

\subsection{Dicke Superradiance}
Dicke superradiance \cite{dicke1954coherence} is a fascinating phenomenon in which a group of excited atoms, as they release their energy, do so collectively and in a coordinated manner. This synchronized decay causes them to behave like a single, powerful antenna that emits a burst of photons. To achieve this effect, the atoms must be closely spaced, with an inter-atomic distance $(d)$ smaller than the wavelength ($\lambda$) of the common electromagnetic mode of their interaction.

When the atoms are sufficiently close together ($d<< d_{critical}$ where $d_{critical} \gtrsim \lambda$), their interaction leads to synchronous emission, where the phase of oscillations of the individual atomic dipoles becomes locked as the emission event progresses in real time. This synchronization enhances the emission rate initially, resulting in a more rapid release of energy compared to the case when atoms are spaced further apart ($d \gtrsim d_{critical}$). In the latter case the spontaneous emission rate decays monotonically in time and is simply proportional to the number of emitters. 

We follow the work by S. Masson et al. \cite{masson2022universality} to simulate the Dicke superradiance and calculate the emission rate by developing an anstaz used for UAVQD scheme. We use the following Hamiltonian for $n$ atomic emitters
\begin{equation}\label{Ham_Dicke}   H=\hbar\sum^n_{i=1}\omega_0\hat{\sigma}^i_{ee}+\hbar\sum^n_{i,j=1}J^{ij}\hat{\sigma}^i_{eg}\hat{\sigma}^j_{ge}
\end{equation}
where $\sigma^i_{ee} = |e\rangle \langle e|_i$ i.e. signifying projector onto the excited state of the $i$-th qubit. Similarly $\sigma^i_{ge}=|g\rangle \langle e|_i$ is a ladder operator for $|g\rangle \to |e\rangle$ at the $i$-th site. 
The coherent interaction matrix elements $J^{ij}$ between the atomic sites in Eq.\ref{Ham_Dicke} can be obtained as
\begin{equation}
    J^{ij}=-Re(\frac{\mu_0\omega_0^2}{\hbar}\mathcal{P}^{\ast}   \cdot \mathcal{G}_0(\mathbf{r}_i,\mathbf{r}_i,\omega_0) \cdot\mathcal{P})
\end{equation}
where $\mathcal{P}$ is the dipole matrix element associated with the transition and $\mathcal{G}_0(\mathbf{r}_i,\mathbf{r}_i,\omega_0)$ is the Green's function often called the propagator of the electromagnetic field between different atoms at positions $\mathbf{r}_i$ and $\mathbf{r}_j$
\begin{equation}
    \mathcal{G}_0(\mathbf{r}_{ij},\omega_0) =\frac{e^{\mathrm{i}k_0r_{ij}}}{4\pi k_0^2r_{ij}^3 }\left[ (k_0^2r_{ij}^2+\mathrm{i}k_0r_{ij}-1)\mathbb{I}+ (-k_0^2r_{ij}^2-3\mathrm{i}k_0r_{ij}+3)\frac{\mathbf{r}_{ij}\otimes\mathbf{r}_{ij}}{r_{ij}^2} \right].
\end{equation}
\noindent where $k_0=2\pi/\lambda_0$, $r_{ij} = r_i - r_j$ and $r_{ij} = \lvert r_{ij} \rvert$. We chose $\omega_0=2\pi$ for all computations henceforth.

\begin{figure}
  \centering
  \includegraphics[width=0.7\textwidth]{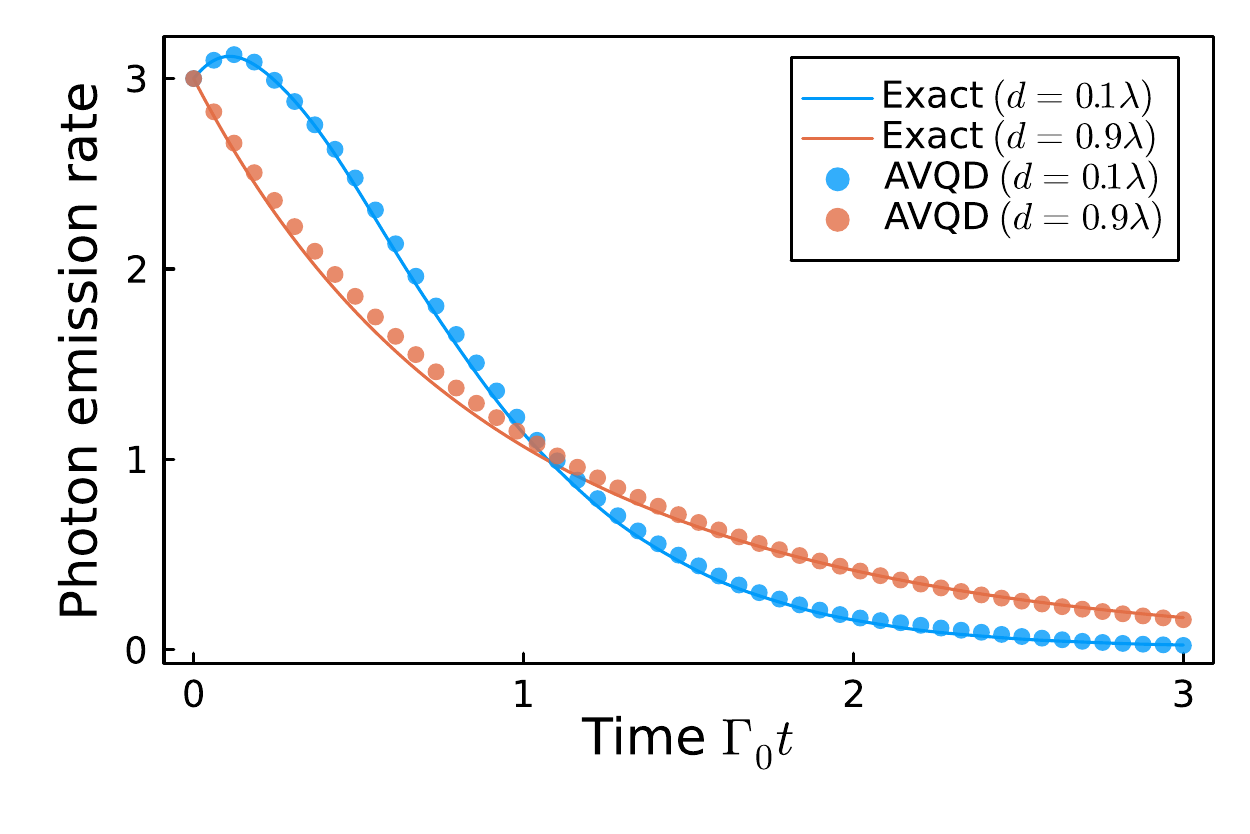}
  \caption{Photon emission rate for the three atom chain configuration is plotted against the time. The solid curves and seperated dots represent the exact and UAVQDS-vectorization calculations respectively for two possible scenarios: when atoms are densely packed i.e. $d=0.1\lambda$ and when atoms are separated at an inter-atomic distance of $d=0.9\lambda$ which is similar to the magnitude of the excitation wavelength of the coupling field. The method was employed with a timestep of $10^{-3}$ time unit, utilizing operator pool P2 and an adaptive threshold of $10^{-1}$.}
  \label{fig:Dicke_data}
\end{figure}

To define jump operators we use a different set of matrix elements $\Gamma^{ij}$ which can be obtained as 
\begin{equation}\label{Gamma_mat_Dicke}
\Gamma^{ij} = -2Im(\frac{\mu_0\omega_0^2}{\hbar}\mathcal{P}^{\ast}   \cdot \mathcal{G}_0(\mathbf{r}_i,\mathbf{r}_i,\omega_0) \cdot\mathcal{P})
\end{equation}
Elements in Eq.\ref{Gamma_mat_Dicke} are thereafter assembled into a matrix $\tilde{\Gamma}$ and diagonalized to obtain eigenvalues as decay rates $\{\Gamma_\nu\}_{\nu=1, N}$. Corresponding eigenvectors $(\alpha_{\nu,1},\alpha_{\nu,1}....\alpha_{\nu,N})^T $  are used to construct collective decay channels/jump operators as follows 

\begin{equation}\label{Jump_op_Dicke}    \hat{L}_{\nu}=\sum^N_{i=1}\alpha_{\nu,i}\hat{\sigma}^i_{ge},\text{ where } \sum^N_{i=1}\alpha_{\nu,i}^{\ast}\alpha_{\mu,i}=\delta_{\nu\mu}\text{ and } \sum^N_{\nu=1}\Gamma_{\nu}|\alpha_{\nu,i}|^2=\Gamma_0 \:\: \forall\:\: i.
\end{equation}
In the above expression $\Gamma_0$ is free space emission rate.
Equipped with Eq.\ref{Ham_Dicke} and Eq.\ref{Jump_op_Dicke} above we can define, the evolution equation for 
density matrix $\rho$ over time t as
\begin{equation}
    \dot{\rho} = -\frac{\mathrm{i}}{\hbar}[H,\rho] +\sum^N_{\nu=1}\frac{\Gamma_{\nu}}{2}\left( 2\hat{L}_{\nu}\rho \hat{L}_{\nu}^{\dagger}-\rho \hat{L}_{\nu}^{\dagger}\hat{L}_{\nu} -\hat{L}_{\nu}^{\dagger}\hat{L}_{\nu}\rho  \right) 
\end{equation}


We selected a linear chain comprising of three atomic emitters and utilized the quantum optics package in Julia\cite{Julia-2017} to obtain the jump operators and decay rates for this configuration.\cite{kramer2018quantumoptics} With these values, we conducted simulations using the vectorization and UAVQDS scheme. We use for illustration two different lattice spacings ($d=0.1\lambda,0.9\lambda$) where $\lambda_0 \propto \frac{1}{\omega_0}$. Our simulation requires 6 qubits in total. We employ the same form of the ansatz as that in the FMO complex as shown in Fig.\ref{fig:FMO_ansatz} and we start the simulation in $|eee\rangle$. It must be emphasized, that an example like Dicke superradiance is extremely difficult to be treated with the trajectory method \cite{chen2024adaptive} as the dynamical evolution doesn't restrict the state space to the single-excitation manifold and the ground state (unlike in FMO). On the contrary due to successive emission, the state can collapse into the any of the manifold of states with lower overall excitation number (or to any superposition of states thereof). This renders the discontinuous update of the parameters of the quantum circuit in the trajectory method difficult which is quintessential to capture collapse and ensure proper execution of the algorithm. In the vectorization protocol such discontinuous updates are completely precluded in favor of only continuous updates as characterized by Eq.\ref{eq:Mtheta_dot=V}. 


In Fig.\ref{fig:Dicke_data} we present the total photon emission rate defined as $\eta (t)= \langle \sum_{\nu=1}^N \Gamma_\nu L_\nu^\dagger L_\nu \rangle(t)$. As has been shown in Ref.\cite{masson2022universality}, we see when the lattice spacing $d = 0.1\lambda \:\:(d << d_{\rm{critical}})$ , $\eta$ when plotted against time $(\Gamma_0 t)$ shows a non-monotonic trend characterized by an initial increase followed by an exponentially decaying tail. The peak maxima in $\eta$ vs $\Gamma_0 t$ curve scales as $\propto O(N^2)$. This is in sharp contrast to the other limit when the lattice spacing is $d=0.9\lambda \:\:(d \approx d_{\rm{critical}})$ when $\eta$ vs $\Gamma_0 t$ curve registers a declining exponential with a decay-constant $\propto O(N)$. We see UAVQD procedure can successfully distinguish both the limits and yield results in good quantitative agreements with the exact ones.

\section{Conclusions}
\label{sec:conclusion}

We have demonstrated the versatility of the UAVQD vectorization method through three different examples, showcasing its applicability to a wide range of open quantum system scenarios. Particularly in the case of Dicke Superradiance, where the state can evolve in many possible pathways because of successive emissions, which makes it difficult to use the other flavour of UAVQD, trajectory method, where system collapses to particular state for measurement. Therefore, we have shown here successfully that the vectorization method is best suited for such continuous update requirement. We have developed a UAVQD operator pool capable of handling the complexity of multiqubit systems. The runtime complexity of our method is significantly lower, i.e. polynomially scaled compared to the exponential scaling of classical methods. In the Appendix we have shown that we can improve this to linear scaling with the scheme which can require only $O(\log_2 N)$ gates, given all-to-all connectivity in qubits. The trapped ion quantum computers have been proven to be promising to handle such ansatz constructions.\cite{linke2017experimental} While demonstrating the successful execution of UAVQD method, we would like to reemphasize one of the shortcomings of UAVQD method is the necessity to choose the correct operator pool, which can be challenging at times. There is room for improvement in our scheme for the decomposition of the ansatz, as well as in the selection of the operator pool. Future work could focus on refining these aspects and studying more examples, such as Dicke Superradiance, where continuous updates are a requirement.

\section*{Acknowledgement}
We acknowledge the financial support of  the National Science Foundation under award number 2124511, CCI Phase I: NSF Center for Quantum Dynamics on Modular Quantum Devices (CQD-MQD). We would like to thank Dr. Wibe Albert de Jong and his research group for sharing the code with us. 
\newpage
\bibliographystyle{MSP}
\bibliography{References}

\begin{thebibliography}{10}
\providecommand{\url}[1]{\texttt{#1}}
\providecommand{\urlprefix}{URL }

\bibitem{breuer2002theory}
H.-P. Breuer, F.~Petruccione,
\newblock \emph{The theory of open quantum systems},
\newblock Oxford University Press, USA, \textbf{2002}.

\bibitem{kais2014introduction}
S.~Kais,
\newblock \emph{Quantum Information and Computation for Chemistry} \textbf{2014}, 1--38.

\bibitem{hu2020quantum}
Z.~Hu, R.~Xia, S.~Kais,
\newblock \emph{Scientific reports} \textbf{2020}, \emph{10}, 1 1.

\bibitem{wang2023simulating}
Y.~Wang, E.~Mulvihill, Z.~Hu, N.~Lyu, S.~Shivpuje, Y.~Liu, M.~B. Soley, E.~Geva, V.~S. Batista, S.~Kais,
\newblock \emph{Journal of Chemical Theory and Computation} \textbf{2023}.

\bibitem{sension2007quantum}
R.~J. Sension,
\newblock \emph{Nature} \textbf{2007}, \emph{446}, 7137 740.

\bibitem{hu2022general}
Z.~Hu, K.~Head-Marsden, D.~A. Mazziotti, P.~Narang, S.~Kais,
\newblock \emph{Quantum} \textbf{2022}, \emph{6} 726.

\bibitem{yeh2012population}
S.-H. Yeh, J.~Zhu, S.~Kais,
\newblock \emph{The Journal of Chemical Physics} \textbf{2012}, \emph{137}, 8.

\bibitem{hu2018double}
Z.~Hu, G.~S. Engel, S.~Kais,
\newblock \emph{Physical Chemistry Chemical Physics} \textbf{2018}, \emph{20}, 47 30032.

\bibitem{you2017plasmon}
X.~You, S.~Ramakrishna, T.~Seideman,
\newblock \emph{ACS Photonics} \textbf{2017}, \emph{4}, 5 1178.

\bibitem{candiotto2017charge}
G.~Candiotto, A.~Torres, K.~T. Mazon, L.~G. Rego,
\newblock \emph{The Journal of Physical Chemistry C} \textbf{2017}, \emph{121}, 42 23276.

\bibitem{hu2021vibration}
Z.~Hu, Z.~Xu, G.~Chen,
\newblock \emph{The Journal of Chemical Physics} \textbf{2021}, \emph{154}, 15.

\bibitem{hu2018dark}
Z.~Hu, G.~S. Engel, F.~H. Alharbi, S.~Kais,
\newblock \emph{The Journal of chemical physics} \textbf{2018}, \emph{148}, 6.

\bibitem{roy2021enhancement}
S.~Roy, Z.~Hu, S.~Kais, P.~Bermel,
\newblock \emph{Advanced Functional Materials} \textbf{2021}, \emph{31}, 23 2100387.

\bibitem{10.1063/1.5096173}
R.~Kosloff,
\newblock \emph{The Journal of Chemical Physics} \textbf{2019}, \emph{150}, 20 204105.

\bibitem{PhysRevB.101.184510}
M.~Majland, K.~S. Christensen, N.~T. Zinner,
\newblock \emph{Phys. Rev. B} \textbf{2020}, \emph{101} 184510.

\bibitem{PhysRevX.4.011015}
N.~L\"orch, J.~Qian, A.~Clerk, F.~Marquardt, K.~Hammerer,
\newblock \emph{Phys. Rev. X} \textbf{2014}, \emph{4} 011015.

\bibitem{PhysRevLett.129.063601}
S.~B. J\"ager, T.~Schmit, G.~Morigi, M.~J. Holland, R.~Betzholz,
\newblock \emph{Phys. Rev. Lett.} \textbf{2022}, \emph{129} 063601.

\bibitem{braaten2017lindblad}
E.~Braaten, H.-W. Hammer, G.~P. Lepage,
\newblock \emph{Physical Review A} \textbf{2017}, \emph{95}, 1 012708.

\bibitem{PhysRevA.82.013640}
R.~A. Pepino, J.~Cooper, D.~Meiser, D.~Z. Anderson, M.~J. Holland,
\newblock \emph{Phys. Rev. A} \textbf{2010}, \emph{82} 013640.

\bibitem{smart2022relaxation}
S.~E. Smart, Z.~Hu, S.~Kais, D.~A. Mazziotti,
\newblock \emph{Communications Physics} \textbf{2022}, \emph{5}, 1 28.

\bibitem{brand2024markovian}
D.~Brand, I.~Sinayskiy, F.~Petruccione,
\newblock \emph{Scientific Reports} \textbf{2024}, \emph{14}, 1 4769.

\bibitem{gorini1978properties}
V.~Gorini, A.~Frigerio, M.~Verri, A.~Kossakowski, E.~Sudarshan,
\newblock \emph{Reports on mathematical physics} \textbf{1978}, \emph{13}, 2 149.

\bibitem{sweke2015universal}
R.~Sweke, I.~Sinayskiy, D.~Bernard, F.~Petruccione,
\newblock \emph{Physical Review A} \textbf{2015}, \emph{91}, 6 062308.

\bibitem{head2021capturing}
K.~Head-Marsden, S.~Krastanov, D.~A. Mazziotti, P.~Narang,
\newblock \emph{Physical Review Research} \textbf{2021}, \emph{3}, 1 013182.

\bibitem{attal2006open}
S.~Attal, A.~Joye, C.-A. Pillet,
\newblock \emph{Open Quantum Systems I: The Hamiltonian Approach},
\newblock Springer, \textbf{2006}.

\bibitem{merkli2022dynamics}
M.~Merkli,
\newblock \emph{Quantum} \textbf{2022}, \emph{6} 616.

\bibitem{lindblad1976generators}
G.~Lindblad,
\newblock \emph{Communications in Mathematical Physics} \textbf{1976}, \emph{48} 119.

\bibitem{kossakowski1972quantum}
A.~Kossakowski,
\newblock \emph{Reports on Mathematical Physics} \textbf{1972}, \emph{3}, 4 247.

\bibitem{arute2019quantum}
F.~Arute, K.~Arya, R.~Babbush, D.~Bacon, J.~C. Bardin, R.~Barends, R.~Biswas, S.~Boixo, F.~G. Brandao, D.~A. Buell, et~al.,
\newblock \emph{Nature} \textbf{2019}, \emph{574}, 7779 505.

\bibitem{lund2017quantum}
A.~P. Lund, M.~J. Bremner, T.~C. Ralph,
\newblock \emph{npj Quantum Information} \textbf{2017}, \emph{3}, 1 15.

\bibitem{sajjan2022quantum}
M.~Sajjan, J.~Li, R.~Selvarajan, S.~H. Sureshbabu, S.~S. Kale, R.~Gupta, V.~Singh, S.~Kais,
\newblock \emph{Chemical Society Reviews} \textbf{2022}.

\bibitem{sajjan2023imaginary}
M.~Sajjan, V.~Singh, R.~Selvarajan, S.~Kais,
\newblock \emph{Physical Review Research} \textbf{2023}, \emph{5}, 1 013146.

\bibitem{li2023toward}
J.~Li, B.~A. Jones, S.~Kais,
\newblock \emph{Science Advances} \textbf{2023}, \emph{9}, 19 eadg4576.

\bibitem{shor1999polynomial}
P.~W. Shor,
\newblock \emph{SIAM review} \textbf{1999}, \emph{41}, 2 303.

\bibitem{aspuru2005simulated}
A.~Aspuru-Guzik, A.~D. Dutoi, P.~J. Love, M.~Head-Gordon,
\newblock \emph{Science} \textbf{2005}, \emph{309}, 5741 1704.

\bibitem{han2021experimental}
J.~Han, W.~Cai, L.~Hu, X.~Mu, Y.~Ma, Y.~Xu, W.~Wang, H.~Wang, Y.~Song, C.-L. Zou, et~al.,
\newblock \emph{Physical Review Letters} \textbf{2021}, \emph{127}, 2 020504.

\bibitem{wang2011quantum}
H.~Wang, S.~Ashhab, F.~Nori,
\newblock \emph{Physical Review A} \textbf{2011}, \emph{83}, 6 062317.

\bibitem{su2020quantum}
H.-Y. Su, Y.~Li,
\newblock \emph{Physical Review A} \textbf{2020}, \emph{101}, 1 012328.

\bibitem{endo2020variational}
S.~Endo, J.~Sun, Y.~Li, S.~C. Benjamin, X.~Yuan,
\newblock \emph{Physical Review Letters} \textbf{2020}, \emph{125}, 1 010501.

\bibitem{suri2023two}
N.~Suri, J.~Barreto, S.~Hadfield, N.~Wiebe, F.~Wudarski, J.~Marshall,
\newblock \emph{Quantum} \textbf{2023}, \emph{7} 1002.

\bibitem{yao2021adaptive}
Y.-X. Yao, N.~Gomes, F.~Zhang, C.-Z. Wang, K.-M. Ho, T.~Iadecola, P.~P. Orth,
\newblock \emph{PRX Quantum} \textbf{2021}, \emph{2} 030307.

\bibitem{zhang2023quantum}
Y.~Zhang, Z.~Hu, Y.~Wang, S.~Kais,
\newblock \emph{The Journal of Physical Chemistry Letters} \textbf{2023}, \emph{14}, 3 832.

\bibitem{benedetti2021hardware}
M.~Benedetti, M.~Fiorentini, M.~Lubasch,
\newblock \emph{Physical Review Research} \textbf{2021}, \emph{3}, 3 033083.

\bibitem{schlimgen2021quantum}
A.~W. Schlimgen, K.~Head-Marsden, L.~M. Sager, P.~Narang, D.~A. Mazziotti,
\newblock \emph{Physical Review Letters} \textbf{2021}, \emph{127}, 27 270503.

\bibitem{PhysRevLett.127.020504}
J.~Han, W.~Cai, L.~Hu, X.~Mu, Y.~Ma, Y.~Xu, W.~Wang, H.~Wang, Y.~P. Song, C.-L. Zou, L.~Sun,
\newblock \emph{Phys. Rev. Lett.} \textbf{2021}, \emph{127} 020504.

\bibitem{chen2022hamiltonian}
H.~Chen, D.~A. Lidar,
\newblock \emph{Communications Physics} \textbf{2022}, \emph{5}, 1 112.

\bibitem{chen2024adaptive}
H.~Chen, N.~Gomes, S.~Niu, W.~A. de~Jong,
\newblock \emph{Quantum} \textbf{2024}, \emph{8} 1252.

\bibitem{blankenship2021molecular}
R.~E. Blankenship,
\newblock \emph{Molecular mechanisms of photosynthesis},
\newblock John Wiley \& Sons, \textbf{2021}.

\bibitem{scholes2017using}
G.~D. Scholes, G.~R. Fleming, L.~X. Chen, A.~Aspuru-Guzik, A.~Buchleitner, D.~F. Coker, G.~S. Engel, R.~Van~Grondelle, A.~Ishizaki, D.~M. Jonas, et~al.,
\newblock \emph{Nature} \textbf{2017}, \emph{543}, 7647 647.

\bibitem{dicke1954coherence}
R.~H. Dicke,
\newblock \emph{Phys. Rev.} \textbf{1954}, \emph{93}, 1 99.

\bibitem{masson2022universality}
S.~J. Masson, A.~Asenjo-Garcia,
\newblock \emph{Nat. Commun.} \textbf{2022}, \emph{13}, 1 2285.

\bibitem{kramer2016optimized}
S.~Kr{\"a}mer, L.~Ostermann, H.~Ritsch,
\newblock \emph{EPL} \textbf{2016}, \emph{114}, 1 14003.

\bibitem{asenjo2017exponential}
A.~Asenjo-Garcia, M.~Moreno-Cardoner, A.~Albrecht, H.~Kimble, D.~E. Chang,
\newblock \emph{Physical Review X} \textbf{2017}, \emph{7}, 3 031024.

\bibitem{sierra2022dicke}
E.~Sierra, S.~J. Masson, A.~Asenjo-Garcia,
\newblock \emph{Phys. Rev. Res.} \textbf{2022}, \emph{4}, 2 023207.

\bibitem{yip2018quantum}
K.~W. Yip, T.~Albash, D.~A. Lidar,
\newblock \emph{Physical Review A} \textbf{2018}, \emph{97}, 2 022116.

\bibitem{horn1991topics}
R.~A. Horn, C.~R. Johnson,
\newblock \emph{Cambridge University Presss, Cambridge} \textbf{1991}, \emph{37} 39.

\bibitem{kamakari2022digital}
H.~Kamakari, S.-N. Sun, M.~Motta, A.~J. Minnich,
\newblock \emph{PRX Quantum} \textbf{2022}, \emph{3}, 1 010320.

\bibitem{nielsen2010quantum}
M.~A. Nielsen, I.~L. Chuang,
\newblock \emph{Quantum computation and quantum information},
\newblock Cambridge university press, \textbf{2010}.

\bibitem{wilde2013quantum}
M.~M. Wilde,
\newblock \emph{Quantum information theory},
\newblock Cambridge university press, \textbf{2013}.

\bibitem{adolphs2006proteins}
J.~Adolphs, T.~Renger,
\newblock \emph{Biophysical journal} \textbf{2006}, \emph{91}, 8 2778.

\bibitem{skochdopole2011functional}
N.~Skochdopole, D.~A. Mazziotti,
\newblock \emph{The Journal of Physical Chemistry Letters} \textbf{2011}, \emph{2}, 23 2989.

\bibitem{bassham1979photosynthetic}
J.~Bassham,
\newblock Photosynthetic carbon metabolism and related processes; encyclopedia of plant physiology vol. 6,(gibbs, m.; latzko, e., eds), \textbf{1979}.

\bibitem{valleau2017absence}
S.~Valleau, R.~A. Studer, F.~H{\"a}se, C.~Kreisbeck, R.~G. Saer, R.~E. Blankenship, E.~I. Shakhnovich, A.~Aspuru-Guzik,
\newblock \emph{ACS Central Science} \textbf{2017}, \emph{3}, 10 1086.

\bibitem{Julia-2017}
J.~Bezanson, A.~Edelman, S.~Karpinski, V.~B. Shah,
\newblock \emph{SIAM {R}eview} \textbf{2017}, \emph{59}, 1 65.

\bibitem{kramer2018quantumoptics}
S.~Kr{\"a}mer, D.~Plankensteiner, L.~Ostermann, H.~Ritsch,
\newblock \emph{Computer Physics Communications} \textbf{2018}, \emph{227} 109.

\bibitem{linke2017experimental}
N.~M. Linke, D.~Maslov, M.~Roetteler, S.~Debnath, C.~Figgatt, K.~A. Landsman, K.~Wright, C.~Monroe,
\newblock \emph{Proceedings of the National Academy of Sciences} \textbf{2017}, \emph{114}, 13 3305.

\end{thebibliography}
\newpage
\section*{Appendix:}
The utilization of multi-qubit gates within an ansatz necessitates their decomposition into basis gates for implementation on a quantum circuit. This decomposition not only provides insights into the depth and quantity of gates required in a quantum device but also aids in understanding its intricacies. Our proposed decomposition strategy involves the most common CNOT gate and single-qubit Rz gate. This decomposition is linearly scalable for k-qubit gates, representing an efficient approach. One can employ the following scheme, (as detailed in the figure caption below) for decomposing all multi-qubit gates in the ansatz used in Fig.\ref{fig:FMO_ansatz}. Even though we show a generic  example of a $k$- qubit ansatz below, it must be emphasized that application of such multi-qubit gates among arbitrary subset of qubits can be easily accomplished in fully connected quantum computers, such as those on trapped ion-based platforms.\cite{linke2017experimental}, 
\begin{figure}[h!]
  \centering
  \includegraphics[width=\textwidth]{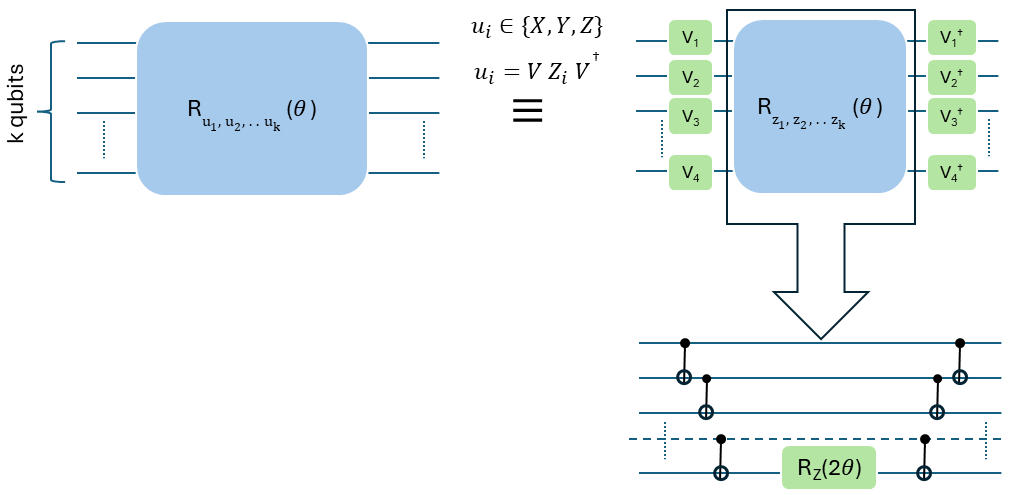}
  \caption{Quantum circuits require the transformation of multi-qubit entangled gates present in the ansatz into CNOT and single-qubit gates. For the general form of the operator present in our ansatz, \(R_{u_1,u_2,u_3,...,u_k}\), we first decompose it using the relation \(u_i = V_iZ_iV_i^\dagger\). Here, \(u_i\) is one of the Pauli operators, i.e., X, Y, and Z, and their corresponding \(V\) are the eigenvectors that diagnolizes the single-qubit unitary into $Z$.(For example for $u_i=X$ the choice of $V_i$=\rm{H}, the conventional Hadamard gate) . In the second step, the \(R_{z_1,z_2,z_3,...,z_k}\) operator in the middle is decomposed into a series of CNOT gates and one \(R_z\) gate. With this strategy, we observe a linear scaling of the number of decomposed gates; that is, for every $k$-qubit entangled gate, we can decompose it into \(O(k)\) basis gates. Such multi-qubit gates can then be implemented across any $k$ qubits in the hardware chosen provided we are using platforms which can successfully prepare all-to-all qubit connectivity.}
  \label{fig:circuit_decomposition}
\end{figure}
\end{document}